\documentclass[runningheads]{llncs}
\usepackage{graphicx}

\usepackage{amssymb}          
\usepackage{multirow}
\usepackage{color}

\begin{document}

\title{GUESR: A Global Unsupervised Data-Enhancement with Bucket-Cluster Sampling for Sequential Recommendation}
\titlerunning{GUESR}
\vspace{-25pt}
\author{Yongqiang Han\inst{1} \and
Likang Wu\inst{1} \and Hao Wang$^{*}$\inst{1} \and Guifeng Wang\inst{2}  \and  Mengdi Zhang\inst{3} \and  Zhi Li\inst{1} \and Defu Lian\inst{1} \and Enhong Chen\inst{1}}
\authorrunning{Y. Han et al.}
%
\institute{Anhui Province Key Laboratory of Big Data Analysis and Application, University of Science and Technology of China, Hefei, China\\ \email{\{harley,wulk,wanghao3,zhili03\}@mail.ustc.edu.cn}, \email{\{liandefu@,cheneh\}@ustc.edu.cn} 
\and HUAWEI TECHNOLOGIES CO., LTD., Hangzhou, China\\ \email{wangguifeng4@huawei.com} 
\and
Meituan-Dianping Group, Beijing, China\\ \email{zhangmengdi02@meituan.com} } 
\maketitle       
 \vspace{-20pt}
\begin{abstract}

Sequential Recommendation is a widely studied paradigm for learning users' dynamic interests from historical interactions for predicting the next potential item.
Although lots of research work has achieved remarkable progress, they are still plagued by the common issues: data sparsity of limited supervised signals and data noise of accidentally clicking. 
To this end, several works have attempted to address these issues, which ignored the complex association of items across several sequences. 
Along this line, with the aim of learning representative item embedding to alleviate this dilemma, we propose GUESR, from the view of graph contrastive learning.  
Specifically, we first construct the Global Item Relationship Graph (GIRG) from all interaction sequences and present the Bucket-Cluster Sampling (BCS) method to conduct the sub-graphs. 
Then, graph contrastive learning on this reduced graph is developed to enhance item representations with complex associations from the global view. 
We subsequently extend the CapsNet module with the elaborately introduced target-attention mechanism to derive users' dynamic preferences. Extensive experimental results have demonstrated our proposed GUESR could not only achieve significant improvements but also could be regarded as a general enhancement strategy to improve the performance in combination with other sequential recommendation methods. 
\vspace{-15pt}

\end{abstract}

\vspace{-25pt}
\section{INTRODUCTION}
\vspace{-10pt}
With the rapid development of the Internet, recommendation systems have been widely employed on online platforms. Among these, sequential recommendation (SR), predicting the next item for users by regarding historical interactions as temporally-ordered sequences, has attracted various attention from both academia and industry. In the recent literature, a large number of works have been proposed and achieved remarkable progress. 
The initial approach is often based on neural networks \cite{gru4rec,caser}.
Considering the different importance of interacted items on the next prediction, an attention mechanism is further introduced to quantify the weights of items in the sequence, such as SASRec \cite{SASRec} and BERT4Rec \cite{bert4rec}.
Some studies explore the adaptation of graph neural networks in SR and capture the complex patterns of items hidden in interaction sequences, such as SR-GNN \cite{srgnn} and GC-SAN \cite{gcsan}. 
However, most of these works still face the problem of data sparsity and noise, which is prone to fail on limited training signals and complex associations between items.   
Meanwhile, contrastive learning techniques have made a great breakthrough in representation learning. Inspired by its success, some methods apply contrastive learning to improve sequential recommendation, such as S\(^3\)Rec \cite{s3rec} and CL4SRec \cite{cl4rec}.

Despite these methods usually achieving remarkable success, there still exist some deficiencies that can be improved. Firstly, most sequential methods exploit the local context of items in each sequence individually, where co-occurrence information between items is sensitive to noise, and the associations that cross several sequences are not well captured. Secondly, the popularity of items approximates a long-tail distribution, and the interactions of many items are very sparse. The representation of these items in existing models may introduce selection bias in some cases. Thirdly, although some recent studies have applied contrastive learning to alleviate the sparsity of interaction data, they usually construct the data augmentation randomly and lack consideration on how to design suitable contrastive learning strategies for the characteristics of sequential recommendation tasks. 

To this end, in this paper, we introduce graph contrastive learning to the sequential recommendation to learn informative representations of items and provide a solid foundation for the portrayal of users' interests accurately. Specifically, we propose a novel framework GUESR, a \textbf{G}lobal \textbf{U}nsupervised Data-\textbf{E}nhancement method for \textbf{S}equential \textbf{R}ecommendation. To be specific, we first construct a Global Item Relationship Graph (GIRG) from all interaction sequences. We quantified different order adjacent information of items as edge weights to obtain complex associations between items and set a threshold to filter out the noise. In this setting, more edges will be removed for items with high popularity, thus reducing item popularity bias. Subsequently, we adopt graph contrastive learning to learn this global association information for learning enhanced item representations. Besides, we present a Bucket-Cluster Sampling (BCS) method to alleviate the sampling bias of improper negatives and uniform the representation space, which takes into account both efficiency and effectiveness. Additionally, we further extend CapsNet \cite{capsulenet} with a target-attention mechanism to derive the users' preferences on multiple interests, which is formulated as the prediction function. Finally, we jointly optimize this paradigm loss and our proposed auxiliary contrastive learning task. To summarize, the contributions of this paper are as follows,
\begin{itemize}

\item We propose a global contrastive data-enhancement framework for the sequential recommendation, termed GUESR, where the graph contrastive learning is adopted on a constructed global graph to capture the complex associations between items across sequences to alleviate the problem of data sparsity and noise. 
\item To alleviate the influence of improper negatives, we present the Bucket Cluster Sampling (BCS) method with consideration of attribute knowledge, which could benefit from both worlds of efficiency and effectiveness.
\item Extensive experiments on publicly available datasets demonstrate that GUESR outperforms state-of-the-art methods. In addition, some analyses further validate that GUESR is a generic module that can improve the performance of other sequentially recommended methods.
\end{itemize}

\vspace{-15pt}
\section{RELATED WORK}
\vspace{-5pt}
\subsection{Sequential Recommendation}
\vspace{-5pt}

Sequential recommendation (SR) aims to predict the next item based on historical interaction sequences. 
With the development of deep learning, several models based on neural networks have been proposed \cite{gru4rec,caser}. Furthermore, the attention mechanism is a powerful tool applied in sequential recommendation, such as SASRec \cite{SASRec} and BERT4Rec \cite{bert4rec}. In recent years, graph neural networks have achieved state-of-the-art performance in processing graph structure data \cite{h1,h2,k1,k2}. 
Since the powerful GNNs can capture complex item transition patterns hidden in user sequences, there are some studies applying GNNs to SR, such as SR-GNN \cite{srgnn} and GC-SAN \cite{gcsan}.
However, most of these methods above are trained by the prediction loss that optimizes the representation of the entire sequence to improve recommendation performance, while ignoring the valuable unsupervised signal.

\vspace{-10pt}
\subsection{Contrastive Learning for Recommendation}
\vspace{-5pt}
Contrastive learning is an emerging unsupervised learning paradigm that has been successfully applied to computer vision and natural language processing. Meanwhile, some models are applying contrastive learning techniques in sequential recommendation scenarios~\cite{s3rec,cl4rec,icl4sr,clcs}, S\(^3\)Rec \cite{s3rec} devises four auxiliary self-supervised objectives for data representation learning by using mutual information maximization. CL4SRec \cite{cl4rec} applies three data augmentation to generate positive pairs, and contrasts positive pairs for learning robust sequential transition patterns.
Despite the achievement, the contrastive learning-based methods for SR mainly focus on learning the self-supervised signals from each sequence. Due to the limited information in a separate sequence, the obtained self-supervised signal is too weak to encode informative embedding.
\vspace{-10pt}
\section{PRELIMINARY}
\vspace{-10pt}
Considering a set of users $U(|U| = M)$ and items $V(|V| = N)$, each user $u \in U$ has a sequence of interacted items sorted in chronological order $I_{u}=[I_{1}^{u}, \ldots, I_{t}^{u}, \ldots, I_{m}^{u}]$ where $m$ is predefined maximum capacity of interacted items and $I_{t}^{u}$ is the item interacted at step $t$.  Given an observed sequence $I_{u}$, the typical task of sequential recommendation is to predict the next item $I_{m+1}^u$ that the user $u$ is most likely to be interacted with.

\begin{figure}
\centering
\includegraphics[scale=0.3]{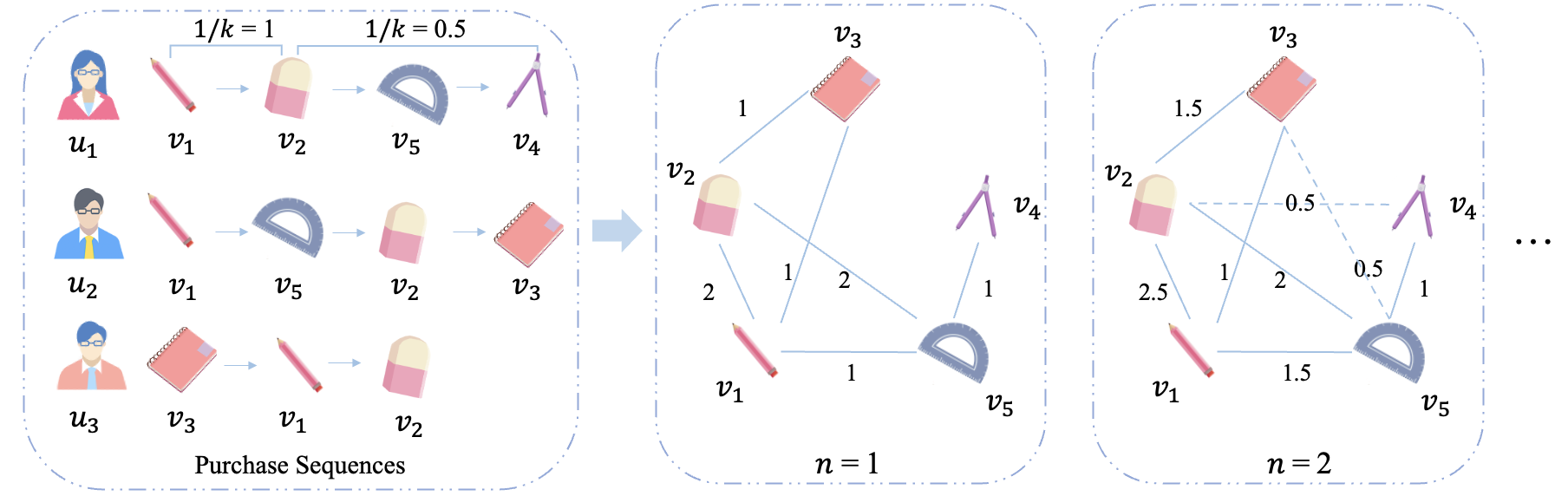}
\caption{An example shows the GIRG construction procedure from item sequences.} \label{fig1}
\end{figure}

\vspace{-15pt}

Unlike existing methods \cite{gru4rec,SASRec,s3rec,cl4rec} that individually exploit the local context of items in each sequence, we first generate a Global Item Relationship Graph (GIRG) \(G\) from all interaction sequences, where each item is connected by an undirected edge. It is worth mentioning that we do not condense repeated items in a sequence. The weight of the edge depends on the adjacency of each item in sequences. Specifically, we define a \(n\)-GIRG where the weight of two edges is the sum of the number of \(k_{th}\) (\(k=[1,2,\cdots,n]\)) direct connections in all sequences and \(k\) represents the adjacent interval of items. This empirical setting is inspired by the success of previous work \cite{lightgcn}. For \((v_i, v_j)\), we calculate $w_{ij}$ and then normalize the edge weights to obtain ${w}^{\prime}_{ij}$. The above process is as follows:

\vspace{-15pt}
\begin{equation} 
w_{ij}= \sum_{u \in  U} \sum_{k=1}^{n} \frac {\delta(|loc(v_{i})-loc(v_{j})|=k \mid v_{i},v_{j} \in I_u )}{k} , 
w^{\prime}_{ij} = Norm(w_{ij}).
\end{equation}
\vspace{-15pt}

Among them, $loc(\cdot)$ represents the position of the item $v_i \in S_u$,  $\delta$ represents the number of times the positions of the two items differ by $k$ in the sequence, and $deg(\cdot)$ denotes the degree of nodes in $G$. After that, we set a hyperparameter threshold $\varepsilon$ to delete some edges~\cite{srwgnn} for reducing noise and get final edge weight $\hat{w}_{ij}$. That is, it mitigates the items' popularity bias since items with high popularity will have more edges removed.

\begin{figure}[t]
   \centering 
    	\includegraphics[width=0.8\textwidth]{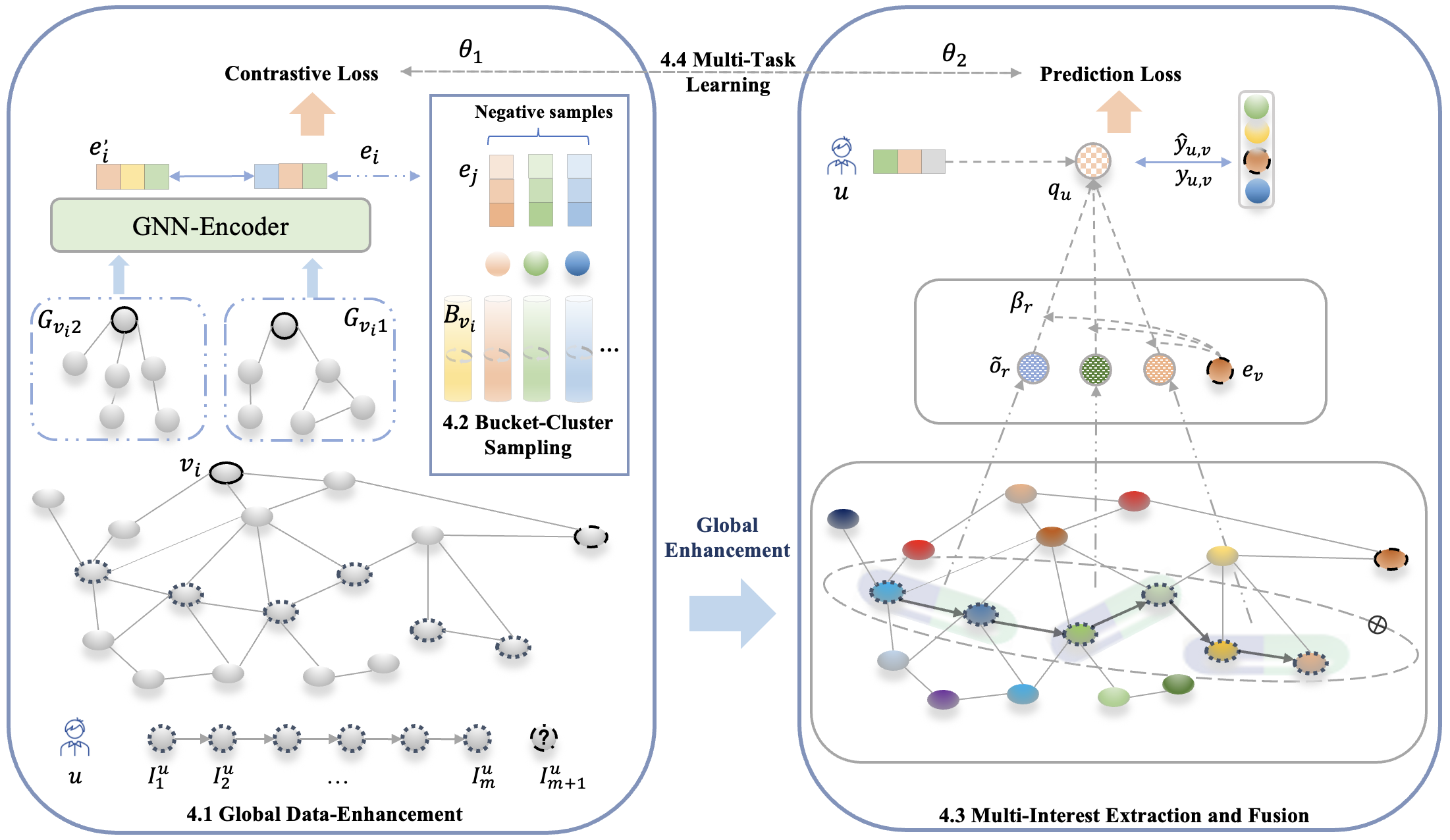}
  	\caption{The network architecture of our proposed GUESR.} \vspace{-12pt}
    \label{fig:framework}
\end{figure}

\vspace{-10pt}
\section{Methodology}
\vspace{-5pt}

\subsection{Global Data-Enhancement}
\vspace{-2pt}

As we know that GCL method is highly dependent on the choice of augmentation scheme \cite{Augmentation}. Based on the exploration of previous work, we adopt a probability-based data augmentation method suitable for global graph scenarios. 
Based on the previous steps, we get the GIRG $G$ for all sequences with the hyperparameters $\varepsilon$ and $n$. After that, for a particular item $v_i$, we sample its neighbor $v_j$ in $G$ by probability $p_{ij}$, which is the ratio of edge weight $\hat{w}_{ij}$ and sampling depth $D$. Please note that the sampling process is iterative, when a node is collected, its adjacent nodes will be put into the sampling pool. 
In this way, the semantic information of the target node can be preserved. 
Through two sampling processes, we obtain two augmented views $G_{v_i1}$ and $G_{v_i2}$ for a particular item $v_i$. Then, the LightGCN~\cite{lightgcn} method is applied as the encoder with the shared parameters in different views. Taking the obtained view $G_{{v_i}1}$ as an example, the information propagation and aggregation at the $l$-th layer of item $v_i$ are as follows:

\vspace{-10pt}
\begin{equation}
e_{i}^{(l+1)}=\sum_{v_j \in N_{i}} \frac{e_{j}^{(l)}}{\sqrt{deg(v_{i})} \sqrt{ deg(v_{j})}},  \qquad e_{i}=\sum_{l=0}^{L} \alpha_{l} e_{i}^{(l)},
\end{equation}
\vspace{-10pt}

where $e_{i}^{(l)}$ denotes the representation of $v_i$ in the $l$-th layer. $N_i$ denotes the set of $v_i$'s neighbors in $G_{v_{i}1}$ and $\alpha_l$ represents the weight in $l$-th layer for the final embedding, which is trained as a model parameter. After $L$ layers of information propagation on  $G_{v_i1}$, we denote the embedding of $v_i$ as $e_i$. Similarly, the embedding of $v_i$ in another view $G_{v_i2}$ is denoted as $e_i^{\prime}$. 
\vspace{-10pt}
\subsection{Bucket-Cluster Sampling}
\vspace{-2pt}
Previous works \cite{cl4rec,s3rec} mostly adopt in-batch negatives or samples from training data at random. Such a way may cause a sampling bias, which will hurt the uniformity of the representation space. To address it, we present a Bucket-Cluster Sampling (BCS) method to alleviate the influence of these improper negatives.

In general, we divide buckets according to the initial attributes and current embeddings iteratively. 
Firstly, we use coarse-grained attribute information provided in the datasets as the basis for bucket splitting. It decides which bucket the item \( v_i \) was originally in. According to the K-means algorithm, the centers of $K$ buckets $(\mu_{1}, \mu_{2}, \ldots, \mu_{K})$ are calculated as the mean value of their item embeddings. Then we calculate the distance between each item embedding $e_i$ and the cluster center $\mu_j$.  Finally, we assign item $v_i$ to the bucket $B_{v_i}$ and uniformly select $N_{neg}$ negative samples in other buckets according to the bucket size. Even though the algorithm is relatively simple, the experiment proves that the sampling method is effective because of the introduction of prior knowledge. The whole process can be formulated as:
\vspace{-5pt}
\begin{equation}
B_{v_i} = \arg\max_{b}\big[(1-\lambda)*\textbf{I}(v_{i} \in B^{orig}_{v_i}) + \lambda*\frac{\left\|e_{i}-\mu_{b}\right\|^{2}}{\sum_{k=1}^{K} \left\|e_{i}-\mu_{k}\right\|^{2}}\big],
\end{equation}
\vspace{-5pt}

here $\textbf{I}(\cdot)$ indicate whether $v_i$ was originally in bucket $B_{v_i}^{orig}$ or not, the hyperparameter $\lambda$ is designed to control the weight of the prior knowledge.

\vspace{-10pt}
\subsection{Multi-Interest Extraction and Fusion}

In this part, we utilize CapsNet \cite{capsulenet} to generate the user’s multiple interests and then conduct the target-attention mechanism to derive the user’s preferences. 
Specifically, In order to utilize critical temporal information for the sequential recommendation, we utilize a Transformer to encode the interaction sequence and obtain the sequential patterns $Z \in \mathbb{R}^{m \times d}$ by additionally introducing a residual operation over $S$ after linear projection with parameter $W^{z} \in \mathbb{R}^{d \times d}$.

Then we define $\textbf{g}=[g_1,g_2,\cdots,g_m]$ as the agreement score, which indicates the relevance of each item towards capsules. Assume that we have $R$ interest capsules, for the $r$-th capsule, its hidden representation $h_r \in \mathbb{R}^{d}$ is summarized over each sequential pattern $z_i \in Z$ by the agreement score with a softmax function. Then, the output of $r$-th capsule denoted as $o_r \in \mathbb{R}^{d}$ is derived from a nonlinear squashing function. 
Immediately after, the agreement score $g_i$ is updated based on the output and the sequential pattern embedding. The above process could be formulated as follows:
\begin{equation}
{h}_{r}=\sum_{i=1}^{m} softmax(g_{i}) {z}_{i},\qquad  o_{r}= \frac{\left\|{h}_{r}\right\|^{2}}{1+\left\|{h}_{r}\right\|^{2}} \frac{{h}_{r}}{\left\|{h}_{r}\right\|},\qquad  
g_{i}=g_{i}+{o}_{r}^{\top} {z}_{i}.
\end{equation}

For each interest capsule, we execute the above process for \(T\) iterations. The output in the final iteration is fed into a fully-connected layer and a ReLU activation function to derive the interest representation:
\begin{equation}
\tilde{o}_{r}={ReLU}\left({o}_{r} {W_r}^{o}\right).
\end{equation}

The weight of interests is affected by the target item \cite{ta}. For example, a sports enthusiast may click on a recommended bike, even after he has clicked several books. Through the above process, we get the interest representations \([\tilde{o}_1 , \tilde{o}_2 \cdots \tilde{o}_R]\), \(R\) is the number of interest capsules.
Given a target item with embedding \(e_{v}\), we  utilize the target-attention mechanism to derive the user preference, the process is as follows:
\begin{equation}
\beta_{r}=\frac{\exp \left(\tilde{o}_{r}^{\top} {e}_{v}\right)}{\sum_{j=1}^{R} \exp \left({\tilde{o}}_{j}^{\top}{e}_{v}\right)}, \qquad  {q}_{u}=\sum_{r=1}^{R} \beta_{r} {\tilde{o}}_{r} + u.
\end{equation}

Among them, \(\beta_j\) is the attention weight and \(q_u\) is the representation of the integrated interest. The user vector \(u\) is added to maintain the uniqueness of the users and the recommendation score is calculated by inner product $\hat{y}_{u, v}={q}_{u} {e}_{v}$. 

\vspace{-10pt}

\subsection{Multi-Task Learning}

With the main prediction task and contrastive learning task, we jointly optimize them in this section. Concretely, we use the InfoNCE Loss~\cite{infonceloss} to distinguish the augmented representations of the same item from others. In addition, the Binary Cross Entropy (BCE) loss is implemented for the prediction task. 
\begin{equation}
\mathcal{L}_{CL}=-\sum_{v_{i} \in V}\left[\log \frac{\exp \left(\mathbf{sim}\left(e_i, e^{\prime}_{i}\right) /\tau\right)  }{ \sum_{j=1}^{N_{neg}} \exp \left(\mathbf{sim}\left(e_i, e_{j}\right)\right/\tau) }\right],
\end{equation}

\vspace{-10pt}

\begin{equation}
\mathcal{L}_{Pred}=-\sum_{u, v}\left[y_{u, v} \ln \left(\hat{y}_{u, v}\right)+\left(1-y_{u, v}\right) \ln \left(1-\hat{y}_{u, v}\right)\right].
\end{equation}
\vspace{-5pt}
Therefore, the final objective function of GUESR is:

\begin{equation}
\mathcal{L}_{Total}=\theta_{1}\mathcal{L}_{Pred}+\theta_{2} \mathcal{L}_{CL} + \theta_{3}\|\Theta\|_{2},
\end{equation}
where \(\theta_3\) is the \(L2 \) regularization parameter to prevent over-fitting.

\vspace{-10pt}
\section{EXPERIMENT}
\vspace{-5pt}

\subsection{Experimental Settings}
\vspace{-5pt}
\textbf{Datasets.}
We conduct experiments on four publicly available datasets in different domains \cite{amazon}. The detailed statistic of datasets is illustrated in Table~\ref{table:Datasets}.

\vspace{-20pt}
\begin{table*}

	\caption{The statistics of datasets.}\vspace{-5pt}
	\centering
	\resizebox{.7\textwidth}{!}{
		\smallskip\begin{tabular}{ccccccc}
			\hline
			\textbf{ Dataset } & \textbf{   \# Users   } & \textbf{   \# Items   } & \textbf{   \# interactions   } & \textbf{   Sparsity   } \\
			\hline
ML-1M    & 6,040   & 3,618   & 836,434        & 96.18\%   \\
Sports   & 35,598  & 18,357  & 296,337        & 99.95\%   \\
Yelp     & 45,478  & 30,709  & 1,777,765      & 99.87\%   \\
Books    & 58,145  & 58,052  & 2,517,437      & 99.93\%   \\ 
			\hline
		\end{tabular}
	}
	\label{table:Datasets}
\end{table*}

\vspace{-20pt}
\hspace{-15pt}\textbf{Implementation Details.}
To start up the experiments, for each user, we select the first 80\% of the interaction sequence as training data, the next 10\% as validation data, and the remained 10\% as the testing data. All Baselines are described in the related work. For a fair comparison, we keep the same experimental environment and search the hyper-parameters carefully. To avoid the sampling bias issues of the candidate selection \cite{fr}, we adopt the full-ranking strategy. To compare performance with state-of-the-art comparison baselines, we adopt two widely used evaluation metrics Recall@K and NDCG@K, and set the top K to 10 and 20. 

\vspace{-15pt}

\subsection{Performance Comparison}
\vspace{-5pt}

In this section, we report the overall recommendation performance by ranking both Recall and NDCG metrics on four public datasets, as shown in Table~\ref{table:performance}, and conclude the following observations. 


\begin{table*}[t]
	\caption{The performance of different models. The best results are in boldface, and the second best results are tagged with the symbol `*' in this paper.}
	\centering
	\resizebox{\textwidth}{!}{
		\smallskip\begin{tabular}{cccccccccccc}
			\hline
			\textbf{Dataset} & \textbf{Metric} & \textbf{BPRMF} & \textbf{GRU4Rec} & \textbf{Caser} & \textbf{SASRec} & \textbf{LightGCN}& \textbf{SR-GNN}& \textbf{GC-SAN}& \textbf{S\(^3\)Rec}& \textbf{CL4Rec} &\textbf{GUESR}\\
			\hline
                \multirow{4}{*}{ML-1M}         & {{R@10}} & 0.1812 & 0.1843  & 0.1756 & 0.1932 & 0.1836   & 0.1885 & 0.1940 & 0.1995 & 0.2034* & \textbf{0.2157}   \\
                         & {{N@10}} & 0.2455 & 0.2465  & 0.2387 & 0.2605 & 0.2463   & 0.2507 & 0.2572 & 0.2638 & 0.2701* & \textbf{0.2879}  \\
                         & {{R@20}} & 0.2712 & 0.2734  & 0.2643 & 0.2895 & 0.2729   & 0.2762 & 0.2850 & 0.2940 & 0.3001* & \textbf{0.3147}   \\
 & N@20 & 0.2558 & 0.2567  & 0.2434 & 0.2730 & 0.2565   & 0.2625 & 0.2707 & 0.2790 & 0.2823* & \textbf{0.2951}   \\ \cline{1-12}
                \multirow{4}{*}{Sports}    & {{R@10}} & 0.1032 & 0.1043  & 0.1021 & 0.1145 & 0.1040   & 0.1132 & 0.1176 & 0.1221& 0.1231* & \textbf{0.1342}  \\
                         & N@10 & 0.0854 & 0.0893  & 0.0833 & 0.0993 & 0.0885   & 0.0935 & 0.0988 & 0.1041 & 0.1054* & \textbf{0.1118}  \\
                         & {{R@20}} & 0.1284 & 0.1302  & 0.1139 & 0.1402 & 0.1298   & 0.1335 & 0.1385 & 0.1438 & 0.1492* & \textbf{0.1579}   \\
 & {{N@20}} & 0.0987 & 0.0992  & 0.0921 & 0.1093 & 0.0991   & 0.1015 & 0.1069 & 0.1123 & 0.1143* & \textbf{0.1239}   \\ \cline{1-12}
                   \multirow{4}{*}{Yelp}       & {{R@10}} & 0.0635 & 0.0643  & 0.0634 & 0.0798 & 0.0647   & 0.0732 & 0.0808 & 0.0884  & 0.0892* & \textbf{0.0981}   \\
                         & {{N@10}} & 0.0452 & 0.0487  & 0.0435 & 0.0601 & 0.0480   & 0.0539 & 0.0610 & 0.0682* & 0.0678  & \textbf{0.0753}  \\
                         & {{R@20}} & 0.1038 & 0.1046  & 0.1042 & 0.1203 & 0.1044   & 0.1236 & 0.1291 & 0.1346  &0.1362* & \textbf{0.1457}  \\
  & {{N@20}} & 0.0572 & 0.0598  & 0.0578 & 0.0710 & 0.0602   & 0.0643 & 0.0728 & 0.0798 & 0.0812* & \textbf{0.0862}  \\ \cline{1-12}
                    \multirow{4}{*}{Books}     & {{R@10}} & 0.0621 & 0.0683  & 0.0624 & 0.0829 & 0.0725   & 0.0763 & 0.0847 & 0.0931* & 0.0921& \textbf{0.1032}   \\
                         & {{N@10}} & 0.0431 & 0.0462  & 0.0445 & 0.0583 & 0.0501   & 0.0506 & 0.0587 & 0.0674* & 0.0672& \textbf{0.0735}  \\
                         & {{R@20}} & 0.0972 & 0.1032  & 0.1001 & 0.1249 & 0.1056   & 0.1116 & 0.1205 & 0.1295 & 0.1307* & \textbf{0.1401}   \\
  & {{N@20}} & 0.0529 & 0.0542  & 0.0546 & 0.0700 & 0.0569   & 0.0609 & 0.0694 & 0.0780    & 0.0795* & \textbf{0.0849}   \\ \cline{1-12}
			\hline
		\end{tabular}
	}\vspace{-10pt}
	\label{table:performance}
\end{table*}

First, we can conclude that SR-GNN and GC-SAN commonly perform better than BPRMF, GRU4Rec, and Caser, which further demonstrates the conclusion of previous work that they can represent more high-order information of users and items. 
Second, we notice that SASRec, which introduced a self-attention mechanism, has achieved better performance than GRU4Rec and Caser, which indicates that self-attention architecture can be suitable for sequence modeling and better capture the long-term dependencies of items in the sequence.
Third, for the self-supervised learning methods, we find that CL4Rec and S\(^3\)Rec consistently perform better than other baselines with single paradigm loss, which demonstrates the effectiveness of introducing self-supervised tasks into sequential recommendation problems. 
Last but not least, we observe that our proposed GUESR consistently performs better than all baselines on both evaluation metrics among all datasets, which indicated the advantage of our proposed global-enriched graph contrastive learning.

\vspace{-15pt}
\subsection{Ablation Study}
\vspace{-5pt}

In this section, we will perform the ablation study to validate and quantify the effectiveness of each component in our proposed GUESR. To be specific, we formulate the following corresponding comparison setting: 1). GUESR-GCL: indicates it removes the graph contrastive learning loss; 2). GUESR-W: represents we create the global item graph without consideration of edge weights and set \(n \)=1 (\(n \) is the max adjacent interval of items.) ; 3). GUESR-BCS: illustrates we use random negative sampling instead of Bucket-Cluster Sampling. The comparison results are presented in Table~\ref{table:ablation}.

\begin{table*}[t]
	\caption{The performance achieved by the different modules of GUESR.}
	\centering
	\resizebox{.9\textwidth}{!}{
		\smallskip\begin{tabular}{cccccccc}
			\hline
			\textbf{Dataset} & \textbf{Metric} & \textbf{GUESR} & \textbf{GUESR-GCL} & \textbf{GUESR-W} & \textbf{GUESR-BCS}  & \textbf{CL4Rec} \\
			\hline
\multicolumn{1}{c}{\multirow{2}{*}{ML-1M}} & \multicolumn{1}{c}{R@20}   & \multicolumn{1}{c}{\textbf{0.3147}} & \multicolumn{1}{c}{0.2802 (-11.0\%)}    & \multicolumn{1}{c}{0.3087 (-1.9\%)}    & \multicolumn{1}{c}{0.2954 (-6.1\%)} & \multicolumn{1}{c}{0.3001*} \\
\multicolumn{1}{c}{}                       & \multicolumn{1}{c}{N@20}   & \multicolumn{1}{c}{\textbf{0.2951}} & \multicolumn{1}{c}{0.2625 (-11.0\%)}    & \multicolumn{1}{c}{0.2886 (-2.2\%)}    & \multicolumn{1}{c}{0.2799 (-5.1\%)} & \multicolumn{1}{c}{0.2823*}\\
\hline
\multicolumn{1}{c}{\multirow{2}{*}{Books}} & \multicolumn{1}{c}{R@20}   & \multicolumn{1}{c}{\textbf{0.1401}} & \multicolumn{1}{c}{0.1149 (-18.0\%)}    & \multicolumn{1}{c}{0.1332 (-5.0\%)}    & \multicolumn{1}{c}{0.1282 (-8.5\%)}  & \multicolumn{1}{c}{0.1307*}\\
\multicolumn{1}{c}{}                       & \multicolumn{1}{c}{N@20}   & \multicolumn{1}{c}{\textbf{0.0849}} & \multicolumn{1}{c}{0.0602 (-29.1\%)}    & \multicolumn{1}{c}{0.0821 (-3.3\%)}    & \multicolumn{1}{c}{0.0723(-14.8\%)} & \multicolumn{1}{c}{0.0795*}\\

			\hline
		\end{tabular}
	}\vspace{-10pt}
	\label{table:ablation}
\end{table*}

From the results shown in Table~\ref{table:ablation}, we can observe the following conclusions: 
Firstly, the performance of GUESR-GCL decreases dramatically on both evaluation metrics in comparison with the original, which proves the effectiveness of employing contrastive learning to mitigate the problems of data sparsity and noise in the sequential recommendation.
Secondly, through comparison with GUESR-W, we can prove that our constructed Global Item Relationship Graph (GIRG) plays an important role in capturing the complex associations of items. 
Thirdly, by comparing with GUESR-BCS, we can conclude that Bucket Cluster Sampling (BCS) method by introducing attribute knowledge could not only alleviate the influence of improper negatives but also can promote efficiency and further improve the effectiveness of our proposed GUESR. 

\vspace{-15pt}

\subsection{Impacts of Enhancement Module}
\vspace{-5pt}
In this part, we will investigate whether our proposed enhancement module in GUESR could be a general framework and flexibly integrate with other sequential recommendation paradigms. Here, we consider GRU4Rec, Caser, and SASRec, and supplement the enhancement module to themselves, which are jointly optimized by both contrastive and prediction loss. From the table, we can observe that all of the enhanced models consistently perform better than the corresponding backbones, which demonstrates the global graph contrastive learning enhancement strategy proposed in GUESR can be a general module directly applied to lots of existing sequential recommendation paradigms. 

\vspace{-10pt}

\begin{table*}
	\caption{The performance comparison with the proposed enhancement module.}
	\centering
	\resizebox{1\textwidth}{!}{
		\smallskip\begin{tabular}{cccccccc}
			\hline
			\textbf{Dataset} & \textbf{Metric} & \textbf{GRU4Rec} & \textbf{EGRU4Rec} & \textbf{Caser} & \textbf{ECaser} & \textbf{SASRec}  & \textbf{ESASRec} \\
			\hline

\multicolumn{1}{c}{\multirow{2}{*}{ML-1M}}  & \multicolumn{1}{c}{R@20}   & \multicolumn{1}{c}{0.2734}  & \multicolumn{1}{c}{0.2809(+2.7\%)}   & \multicolumn{1}{c}{0.2643} & \multicolumn{1}{c}{0.2760 (+4.4\%)} & \multicolumn{1}{c}{0.2895} & \multicolumn{1}{c}{0.2941(+1.6\%)}   \\
\multicolumn{1}{c}{}                        & \multicolumn{1}{c}{N@20}   & \multicolumn{1}{c}{0.2567}  & \multicolumn{1}{c}{0.2635(+2.7\%)}   & \multicolumn{1}{c}{0.2434} & \multicolumn{1}{c}{0.2602 (+7.0\%)} & \multicolumn{1}{c}{0.2730} & \multicolumn{1}{c}{0.2798(+2.5\%)}  \\
\hline
\multicolumn{1}{c}{\multirow{2}{*}{Sports}} & \multicolumn{1}{c}{R@20}   & \multicolumn{1}{c}{0.1302}  & \multicolumn{1}{c}{0.1354(+4.0\%)}   & \multicolumn{1}{c}{0.1139} & \multicolumn{1}{c}{0.1332(+17.0\%)} & \multicolumn{1}{c}{0.1402} & \multicolumn{1}{c}{0.1435(+2.4\%)}   \\
\multicolumn{1}{c}{}                        & \multicolumn{1}{c}{N@20}   & \multicolumn{1}{c}{0.0992}  & \multicolumn{1}{c}{0.1055(+6.4\%)}   & \multicolumn{1}{c}{0.0921} & \multicolumn{1}{c}{0.1021(+10.9\%)} & \multicolumn{1}{c}{0.1093} & \multicolumn{1}{c}{0.1117(+2.2\%)} \\

			\hline
		\end{tabular}
	}
	\label{table:enhancement}
\end{table*}
\vspace{-30pt}

\section{CONCLUSION}
\vspace{-10pt}
In this work, we proposed a novel graph contrastive learning paradigm for the sequential recommendation problem, termed GUESR, to explicitly capture potential relevance within both local and global contexts of items. Extensive experiments on four public datasets demonstrate have demonstrated our proposed GUESR could not only achieve significant improvements but also could be regarded as a general enhancement strategy to improve the performance gains in combination with other sequential recommendation methods. 
\vspace{-10pt}
\section{ACKNOWLEDGEMENTS}
\vspace{-10pt}
This research was supported by grants from the National Natural Science Foundation of China (No. 62202443). This research was also supported by Meituan Group.

\bibliographystyle{splncs04}

\vspace{-10pt}
\bibliography{reference}
\vspace{-35pt}

\end{document}